\pdfoutput=1
\documentclass{article}
\usepackage{graphicx, indentfirst, hyperref, natbib}
\usepackage[a4paper, margin = 3cm]{geometry}
\makeatletter
\let\mailmark\@fnsymbol
\makeatother

\newcommand*{\cf}{\emph{cf.}}
\newcommand*{\eg}{\emph{eg.}}
\newcommand*{\vs}{\emph{vs.}}
\newcommand*{\etc}{\emph{etc}}
\newcommand*{\lb}{\linebreak[1]}
\newcommand*{\figref}[1]{Figure \ref{fig:#1}}
\newcommand*{\ssref}[1]{Subsection \ref{ssec:#1}}
\let\thxmark\textsuperscript
\let\cite\citep

\begin{document}
\title{Better automation of beamline control at HEPS}
\author{%
	Yu Liu\thxmark{1,\mailmark{1}}, Xue-Wei Dong\thxmark{1}, Gang Li\thxmark{1}%
}
\date{}
\maketitle
\begingroup
\renewcommand{\thefootnote}{\fnsymbol{footnote}}
\footnotetext[1]{\ Correspondence e-mail: \texttt{liuyu91@ihep.ac.cn}.}
\endgroup
\footnotetext[1]{\ %
	Institute of High Energy Physics, Chinese Academy of Sciences,
	Beijing 100049, People's Republic of China.%
}

\section*{Synopsis}

Keywords: EPICS; beamline control; package management;
configuration management; software architecture.

A simple, flexible packaging system for EPICS modules and related software is
implemented at HEPS, which also produces reusable modular IOC executables that
can be composed to replace many multi-device IOC applications.  A configuration
convention is suggested that helps to implement easily maintainable multi-IOC
setups; also introduced is our ongoing project of comprehensive beamline
services to further simplify configuration management.

\section*{Abstract}

At the High Energy Photon Source (HEPS) where up to 90 beamlines can be
provided in the future, minimisation of workload for individual beamlines
and maximisation of knowledge about one beamline that can be applied to other
beamlines is essential to minimise the total complexity in beamline control.
Presented in this paper are our efforts to achieve these goals by composing
relatively simple utilities and mechanisms to automate tasks, and always
remembering to keep our automation solutions simple and clear.  After an
introduction to our choice of basic software in EPICS-based beamline control,
the issues encountered in introducing package management to EPICS modules, as
well as our solutions to them, are presented; then the design and implementation
of our packaging system is concisely discussed.  After a presentation of our
efforts to reduce the need for self-built multi-device EPICS IOC applications
by providing reusable modular IOC executables, our implementation of easily
maintainable multi-IOC setups through the separation and minimisation of
each user's IOC configurations is given.  Finally, the ongoing project of
comprehensive beamline services at HEPS to further simplify configuration
management on multiple scales, ranging from individual beamline devices
to all beamlines at HEPS, is introduced.

\section{Introduction}

The High Energy Photon Source (HEPS) \cite{jiao2018} is a 4th-generation
synchrotron radiation facility under steady construction in Beijing, China.
14 beamlines will be provided in 2025, when Phase I of HEPS is planned to open
up to the public; up to about 90 beamlines can be provided in further phases.
The beamline control system at HEPS \cite{chu2018} is mainly responsible for
the proper operation of individual beamline devices and the abstraction of the
various device-specific control interfaces for these devices as consistent
interfaces.  Given the huge number of beamlines, it is undoubted that for
unnecessary work to be reduced, reusable designs and plans are essentials
instead of luxuries.  In this paper, we focus on the beamline control system of
the facility; nevertheless, we believe that in other parts like the accelerator
control system, or even in other large scientific facilities like free-electron
laser facilities and spallation neutron sources, the same principles described
here are still applicable, and most techniques can be adapted.

Given the background above, the tasks of beamline control can be considered
a kind of computer system administration with a focus on diverse device support
and large-scale deployment; with this focus, it is easy to note that tasks for
HEPS beamline control need to be done in simple yet reproducible ways.  For
beamline control at HEPS, we predominantly use Unix-like operating systems,
where off-the-shelf utility programs are abundant and it is very easy to compose
these utilities to automate tasks.  With the utilities, we can express many
tasks as command line code, and further abstract common tasks as scripts; as
the amount of manual operations decrease, the tasks cost less time and labour,
yet the results become more reliable and maintainable -- in other words, both
efficiency and reproducibility increase.

If ``control'' is to be seen as a kind of ``automation'', then the practice
above can be seen as ``automation of automation''; in essence, what we do is
programming specific to the domain of system administration.  When designing
and implementing the beamline control system of HEPS, we find that by seriously
considering the programming nature of tasks, we can often identify the key
factors in problems; we may find surprisingly simple solutions to these
problems by the composition of relatively simple mechanisms and utilities,
like the use of the \emph{flock} utility to implement critical sections
(\cf\ \ssref{cfg}).  In this paper, by showing some examples for the practice
above, we attempt to convey the idea that it is beneficial to regard beamline
control as a serious domain of programming.

As is the case with any other kind of programming, while the abstraction of
beamline control tasks reduces complexity, it also brings about additional
complexity itself.  For multiple reasons, Experimental Physics and Industrial
Control System (EPICS) is chosen as the basis for the device control interfaces
at HEPS.  We find that when performing certain tasks, by deviating from the
``orthodox'' EPICS practice, we are able to make the task much simpler --
sometimes by an order of magnitude, or even more; in some cases, we also deviate
from the ``orthodox'' in certain non-EPICS parts of beamline control.  Therefore
in this paper, what we discuss is not only ``automation of automation'' itself,
but also our choices of strategy when implementing certain requirements,
and more importantly the rationale behind the choices.

\section{Package management}
\subsection{Choice of basic software}\label{ssec:basic}

Due to the requirement of large-scale deployment at HEPS, a first issue to
consider is the installation of supporting software for the many kinds of
devices used in beamline control.  The automation of this task belongs to the
field of package management, which revolves around the automated installation
of software modules and the automatic management of dependencies.  There are
many package managers, like \emph{DPKG}/\emph{APT} of Debian, \emph{RPM}/%
\emph{yum} of Red Hat, \emph{Portage} of Gentoo, \etc; there are also
cross-distribution package managers like \emph{Conda}, \emph{pkgsrc} and
\emph{Guix}, and we are aware of efforts, like \cite{bertrand2019}, to use
this kind of package managers for software involved in beamline control.
However, we find that in order to properly handle dependencies, we need to
duplicate efforts in packaging software already provided by the distribution,
or otherwise we would need to resort to the official package manager of the
distribution for many dependencies.  The latter would not only be inconvenient,
but also degrade the portability of our packages, which is an important reason
to use cross-distribution package managers in the first place.

For various reasons, CentOS is currently the primary operating system running
on computers in control systems at our institute.  For us, while CentOS is
not known for minimalism/simplicity or flexibility, it is usable enough for
automation work in control systems.  More importantly, the strong compatibility
between minor releases (like 7.1 and 7.9) of CentOS makes its basic behaviour
(without deep customisation to the system itself) quite predictable.  Factors
like the inclusion of all official base packages for a minor release in single
\verb|.iso| files (like \texttt{CentOS-7-x86\string_64-Everything-2009\lb.iso})
and the availability of deployment tools like \emph{Kickstart} (\cf\ %
\ssref{cfg}) also contribute to easy automated batch deployment.  Because of
the lifecycle of CentOS releases, we currently base the beamline control system
of HEPS on CentOS 7.  We are aware of the discontinuation of non-rolling
releases of CentOS 8 \cite{bowen2020} and the resulting uncertainty; depending
on future developments of the situation, we will evaluate the migration to a
suitable alternative Linux distribution one or two years prior to the public
opening of HEPS Phase I in 2025.  We estimate that the migration will take
roughly 3 months if we are to use a Red Hat-like distribution, or about 6 months
if not.  Anyway, for reasons above, we currently use \emph{RPM}, the official
packaging format of CentOS, for beamline control software.

We are aware of the trend in the EPICS community to migrate to EPICS 7, and the
declining activity resulted around EPICS 3; considering this background, we plan
to use EPICS 7 whenever applicable, so that we have the best community resources
around the version we use.  However, due to the relatively immature development
and documentation status of new features (most importantly the PVA protocol) in
EPICS 7, we deliberately avoid these features, and instead \emph{use EPICS 7
like EPICS 3}.  Finally, as is quite common with EPICS in beamline control at
major synchrotron radiation facilities, we extensively use the \emph{synApps}
collection \cite{mooney2010} of EPICS modules for HEPS; this has very profound
effects on the packaging of EPICS for HEPS, which is discussed in \ssref{epkg}.
Here we also note that although the official documentation of \emph{synApps}
does not explicitly state about its compatibility with EPICS 7, very few
build-time or runtime errors have actually occurred in our experience
that are traced back to using EPICS 7, which proves that EPICS 7
is mostly backward compatible with EPICS 3 in terms of
functionalities shared between both major versions.

\subsection{Packaging EPICS}\label{ssec:epkg}

EPICS base explicitly supports installation to a separate location with few
negative effects on downstream code like \emph{synApps} \etc, by specifying
the \verb|${INSTALL_LOCATION}| variable which is fairly similar to the
\verb|${DESTDIR}| variable supported by many open-source software projects.
However, on the current version of \emph{synApps} homepage, the recommended way
to install the software collection is to extract the source code archive into a
certain directory (like \texttt{/opt/epics/support}) in the system, build the
source code, and then use the built libraries and executables in place.
So there is no separation between building and installation phases with
\emph{synApps}, and consequently no clear distinction between source code and
files which would be installed separately if \verb|${INSTALL_LOCATION}| was
used.  This allows applications to refer to code that is not explicitly
to be ``installed'', \eg\ \emph{areaDetector} modules referencing files in
\verb|${ADCORE}/ADApp/Db|.  While perhaps convenient, this also results in
\emph{implicit dependence on the synApps directory layout},
which is very unfriendly to packaging.

A first issue we can easily notice is the directory layout.  Ideally, there
should be preset directories for different types of installed files: executables
to a common directory, libraries to another common directory, \etc.  This way,
modules just need to refer to these preset directories for necessary files
without worrying about where to find each and every module they depend on.
There do exist attempts to package EPICS modules in this way, with the NSLS-II
package repository for EPICS and RTEMS \cite{bnl2018} being perhaps the most
well-known example.  Nevertheless as far as we know, none of them provide a
satisfactory coverage of modules we are interested in, most importantly
\emph{motor} and \emph{areaDetector} modules.  From the open-source packaging
code we obtain from these projects, we conclude that attempts to use use common
directories for different modules instead of the \emph{synApps} layout require
an inordinate amount of work by packagers, who need to work around references
outside \verb|${INSTALL_LOCATION}| that may occur in too many ways possible to
handle in a way representable by simple, clear code.  For these reasons, the
\emph{synApps} layout is preserved in our packaging system; and not only modules
provided by \emph{synApps} are organised in its original way, extra modules,
including but not limited to additional motor and area detector modules, are
also put into directories similar to the built-in modules they look like.

The second issue is file conflicts.  With \verb|${INSTALL_LOCATION}| specified,
files in the \verb|configure| directory are copied into the \verb|configure|
subdirectory of \verb|${INSTALL_LOCATION}|; no consideration is taken for the
coexistence of multiple modules, which may have conflicting \verb|configure|
contents.  This kind of conflicts also contribute to the difficulty in proper
packaging of EPICS modules; with the \emph{synApps} directory layout, we do
not need to worry about the conflicts between \verb|configure| contents of
modules, but another kind of file conflicts arise.  Because our system builds
\emph{synApps} and many other EPICS modules in place, we cannot directly
build them on a computer with themselves already installed, which is highly
inconvenient; we used to solve this by running our packaging system in virtual
machines, and now we use Docker containers (\cf\ \ssref{builder}).

The third issue is permissions.  The \emph{synApps} directory layout is
\emph{unrelocatable}: the directories, even as a whole, cannot be moved around
once it has been built, mainly because of parameters like library search paths
and environment variables set to certain absolute paths by the build system
of EPICS.  Because of this, our packaging system builds \emph{synApps} and
other EPICS modules depending on it in \texttt{/opt/epics/support}; to avoid
accidentally tampering with system files, the packaging system itself is
run as a normal user.  However, for the same purpose of avoiding unintended
manipulation of packaged files by users, these files should normally be
installed into a directory hierarchy owned by the \verb|root| user, and
themselves also be owned by \verb|root|.  So in order to build \emph{synApps}
and many other modules, we need write access in directory hierarchies owned by
\verb|root|, even though we build packages as a normal user.  We resolve this
issue by using the \emph{sudo} utility for controlled privilege escalation:
we set up password-less \emph{sudo} permission for a ``builder'' user, and
run our packaging system as this user, so it can acquire \verb|root| privilege
when necessary without manual interaction.  The use of \verb|root| privilege
is purposefully minimised: \eg\ when building \emph{synApps}, we first use
it to move a user-owned \verb|support| hierarchy, extracted from the source
code archive and properly patched, into the \verb|/opt/epics| directory;
we also use \verb|root| privilege to move the mostly-built \verb|support|
hierarchy into a \emph{staging directory} specified by the packaging tool
we use, and then do some necessary processing that can proceed without the
privilege.  For certain modules (like motor modules that are not yet integrated
into \emph{synApps}), some built files may end up in directories (like
\texttt{/opt/epics/support/motor/db}) where files from other modules can
also exist.  We handle this kind of situations by temporarily change the owner
of directories with files from multiple modules to the builder user in an early
phase, and selectively move the non-\verb|root|-owned files in these directories
into corresponding directories inside the staging directory when the building
procedure is mostly done (\cf\ \figref{spec-skel}); the selective moving may be
regarded as a case where the separation between \verb|root| and non-\verb|root|
operations actually simplifies the building procedure.

\subsection{Beyond EPICS itself}\label{ssec:builder}

\begin{figure}[htbp]\centering
\includegraphics[width = 0.86\textwidth]{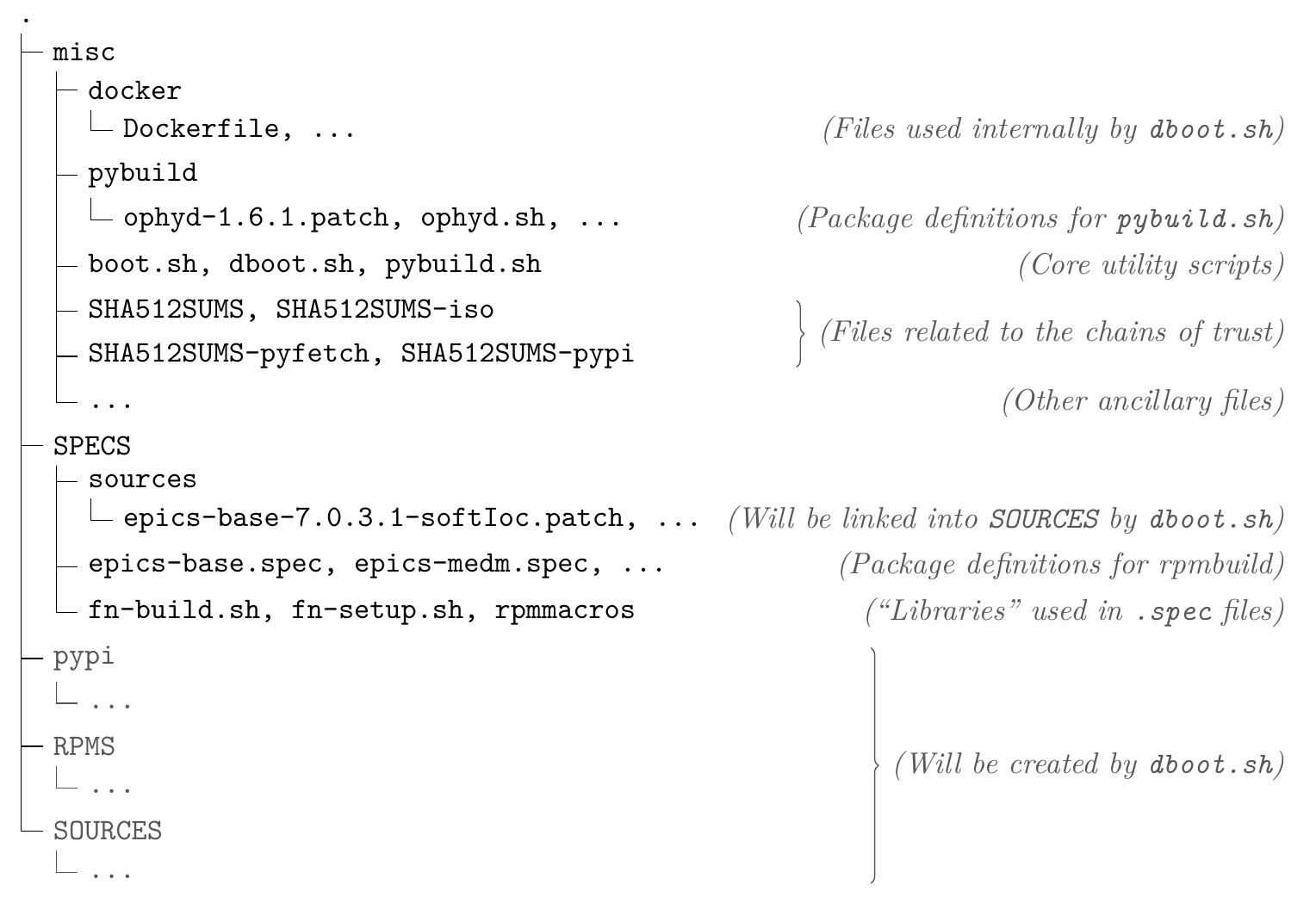}
\caption{Directory layout of our packaging system}\label{fig:builder-skel}
\end{figure}

Considering the limited human resources at our facility, maintainability is a
first concern in the design of our packaging system.  For this reason, we strive
to make the system as ``small'' as reasonable: we keep the packaging code simple
and clear, with proper abstractions that are thin yet flexible enough.  A first
choice we do is about the organisation of packaging code; one approach is to
put the code specific to one package alongside the source code archive(s),
then generate a \emph{source package}, and finally build the binary package we
normally use.  This is how Debian-like and Red Hat-like distributions usually
do, and is also how the NSLS-II package repository is created; however, a most
important drawback of this approach is that the packager may be tempted to apply
site-specific modifications directly in source code repositories.  In a sense,
this disperses packaging code into original upstream code, which creates
difficulty in packaging updates and reviewing modifications; we feel this
is at least an implicit reason for the difficulty in maintenance of the
NSLS-II packages \cite{lange2021}.  Due to reasons above, we put all
our packaging code in a single code repository (\figref{builder-skel};
a fully open-source edition of our packaging system is available
at \url{https://github.com/CasperVector/ihep-pkg-ose}), and directly
build binary packages from there (the source package phase is optional in
CentOS 7, and is also not used in our packaging system); this approach is
also prevalent in more recent Linux distributions like Alpine and Void.

\begin{figure}[htbp]\centering
\includegraphics[width = 0.9\textwidth]{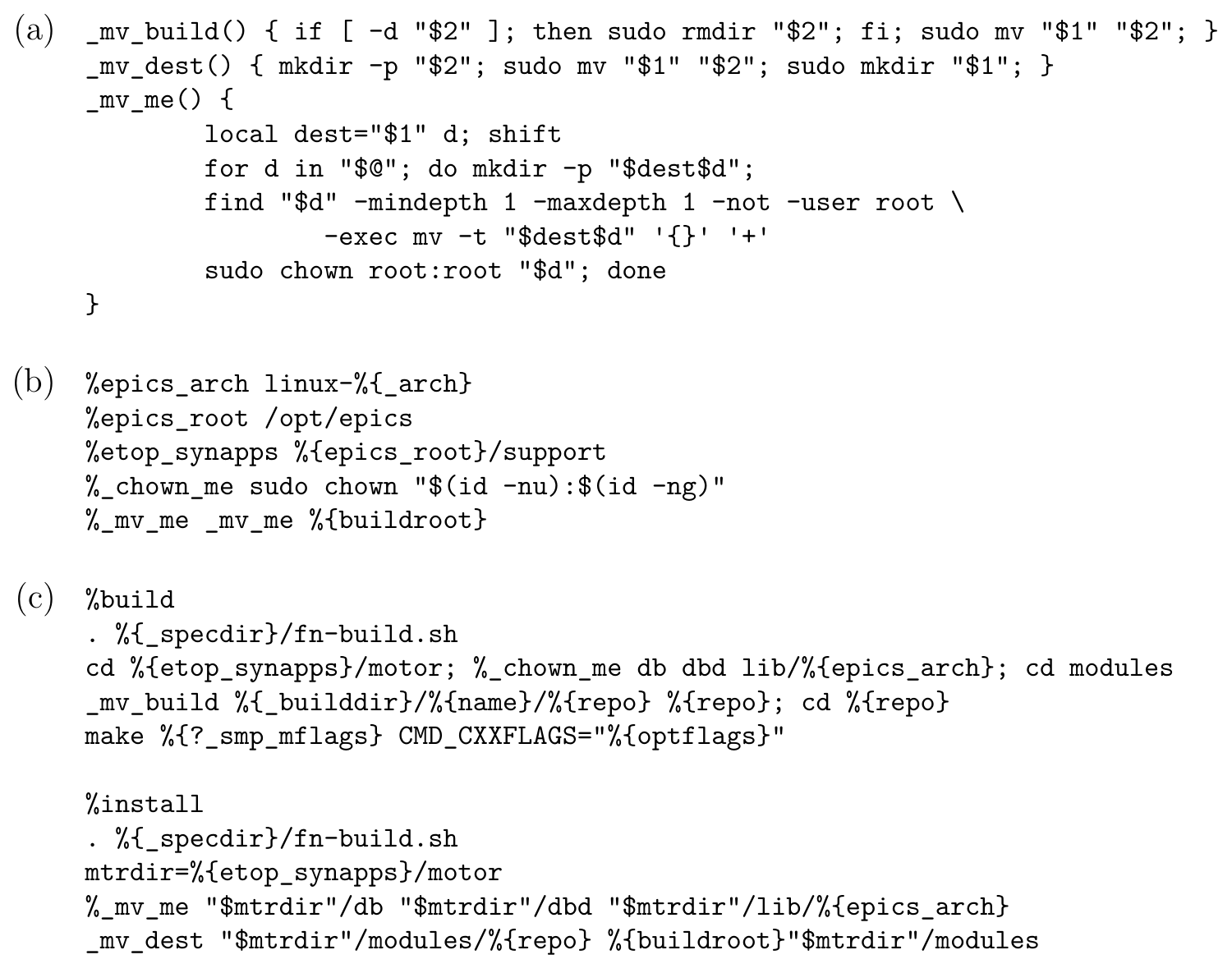}
\caption{%
	Example snippets from (a) \texttt{fn-build.sh}, (b) \texttt{rpmmacros}
	and (c) our \texttt{.spec} files for extra motor modules (where
	\texttt{\%\string{repo\string}} has been set to strings
	like \texttt{motorAcsMotion} earlier)%
}\label{fig:spec-skel}
\end{figure}

\begin{figure}[htbp]\centering
\includegraphics[width = 0.9\textwidth]{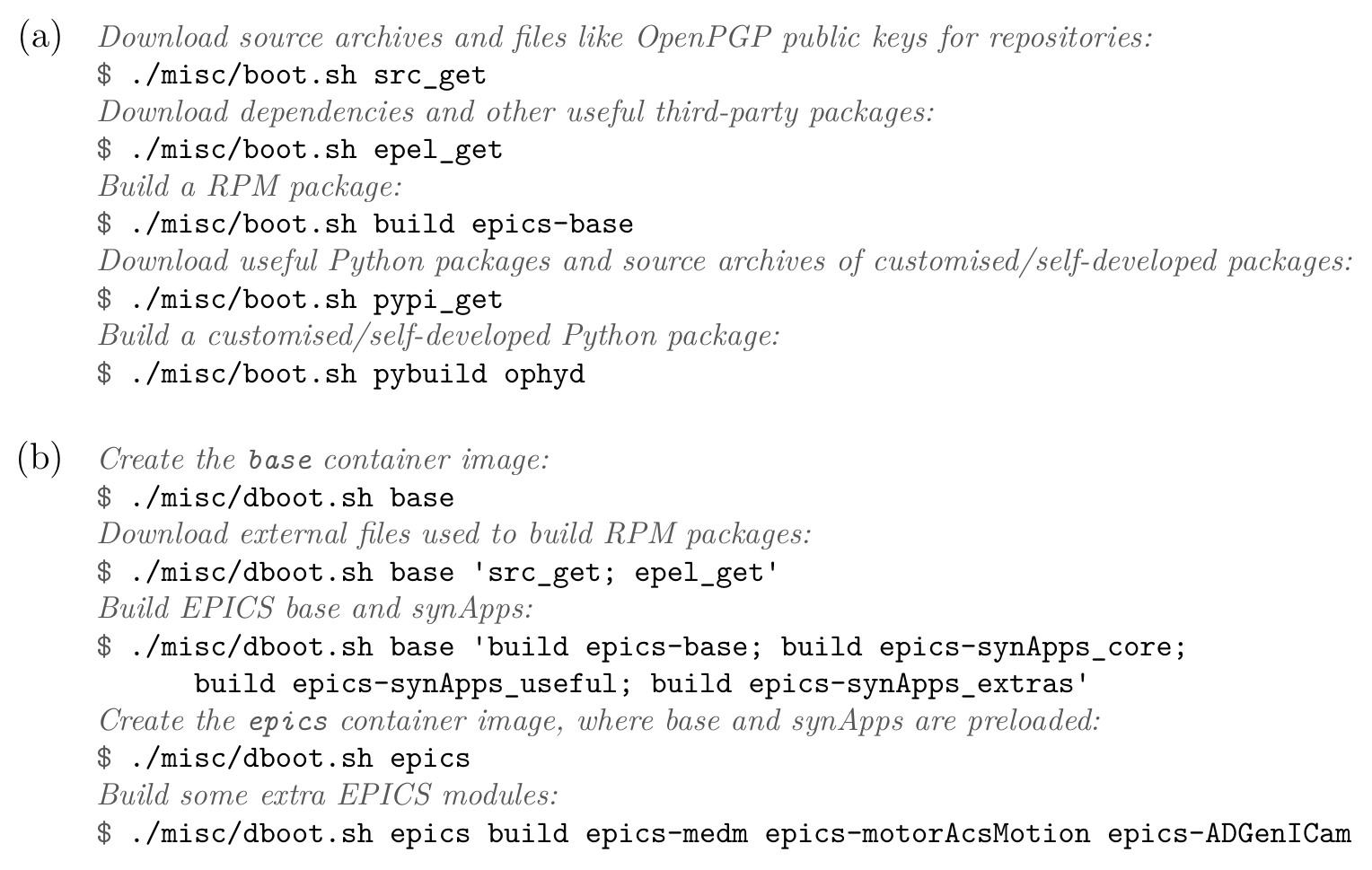}
\caption{%
	Example usage of (a) \texttt{boot.sh} and (b) \texttt{dboot.sh}, both
	expected to be executed from the root directory of the build system%
}\label{fig:boot-usage}
\end{figure}

As has been discussed in \ssref{epkg} and will be further exemplified in this
paper, in order to reduce the intrinsic amount of tasks needed in packaging,
we attempt to package EPICS and related software modules similar to what have
been intended by their authors, with only relatively minor and non-intrusive
modifications.  To reduce duplication in the actual packaging code, multiple
measures have also been taken to abstract common tasks, like how \emph{eclasses}
for Gentoo \emph{ebuild} abstract common tasks.  Because the majority of
packaging code is written in the Shell language, we currently use two library
scripts, \verb|fn-build.sh| and \verb|fn-setup.sh| (\figref{spec-skel}(a)), to
define abstractions for most tasks in the \verb|.spec| package definition files
used in \emph{RPM} packaging; they are not combined into one script because the
code in \verb|fn-build.sh| is only used during the building procedure, while
the code in\verb|fn-setup.sh| needs to be used when the binary package is
installed into the system.  Certain common definitions are needed by both
\verb|.spec| files themselves and the packaging code therein, so we define
them in a \verb|rpmmacros| file (\figref{spec-skel}(b)) that will be loaded by
\emph{rpmbuild}, the official program to build \emph{RPM} packages; also present
in \verb|rpmmacros| are wrapper macros that pass parameters, which are known to
\emph{rpmbuild} but unknown to the library scripts, to functions defined in the
latter, so as to further simplify \verb|.spec| files (\figref{spec-skel}(c)).
By the way, although we follow the \emph{synApps} directory layout, we also try
to keep a slightly higher level of granularity by packaging basic modules (like
\emph{asyn} and \emph{calc}), frequently used modules (like \emph{motor} and
\emph{areaDetector}) and other modules as three separate packages (\cf\ %
\figref{boot-usage}(b)); this not only helps user to slim down systems,
but also helps packagers to better understand the dependencies.

Apart from running \emph{rpmbuild}, quite a lot of ancillary tasks are
also involved in the creation and management of a readily usable repository
of binary packages: downloading and managing source archives, downloading
build-time or runtime dependencies (\eg\ \emph{re2c} required by \emph{synApps})
that are not provided by the base repositories, running \emph{createrepo}
to generate necessary metadata for the repository, \etc.  We use a script,
\verb|boot.sh|, to abstract these tasks, so that the packager only needs to
run simple commands (\figref{boot-usage}(a)) to do common tasks.  A point worth
mentioning is that instead of downloading dependencies as needed when building
a package, we use a locally mounted CentOS \verb|Everything| image (\cf\ %
\ssref{basic}) to provide all base dependencies, and batch-download all
non-base dependencies into the \verb|epel| subdirectory of our repository with
the \verb|epel_get| subcommand of \verb|boot.sh|.  With the same mechanism, we
also download frequently used third-party packages, like \emph{procServ} and
Docker, into our repository, so that our users (which are not very skilled
Linux users in average) have a one-stop solution for common packages.  The
source archives can also be batch-downloaded (using the \verb|src_get|
subcommand); the fact that all source materials can be batch-downloaded
enables us to pre-download the materials, transfer them to a network-isolated
machine with tens of CPU cores, and then perform the building procedure alone
on said machine, thus exploiting high-performance machines
without regular network access for big tasks.

\begin{figure}[htbp]\centering
\includegraphics[width = 0.86\textwidth]{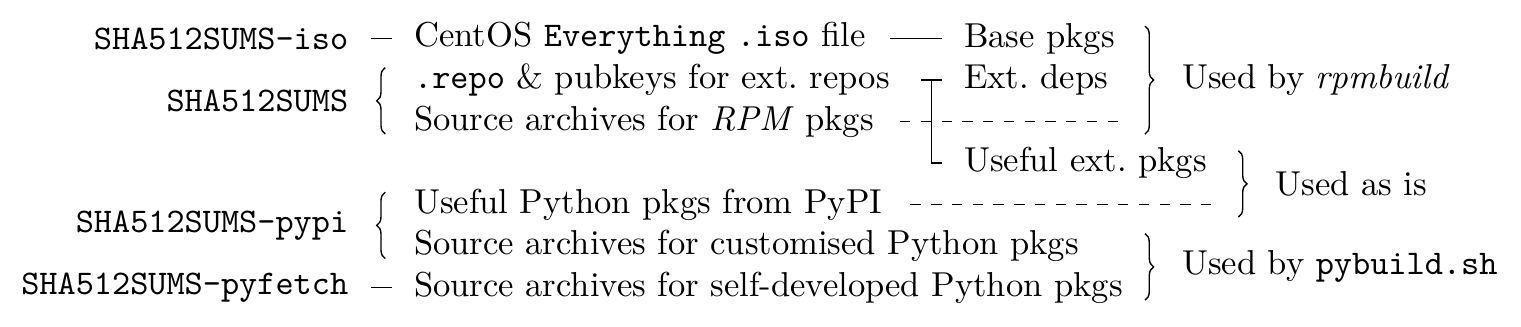}
\caption{Chains of trust for external files in our packaging system}
\label{fig:trust-chain}
\end{figure}

\begin{figure}[htbp]\centering
\includegraphics[width = 0.9\textwidth]{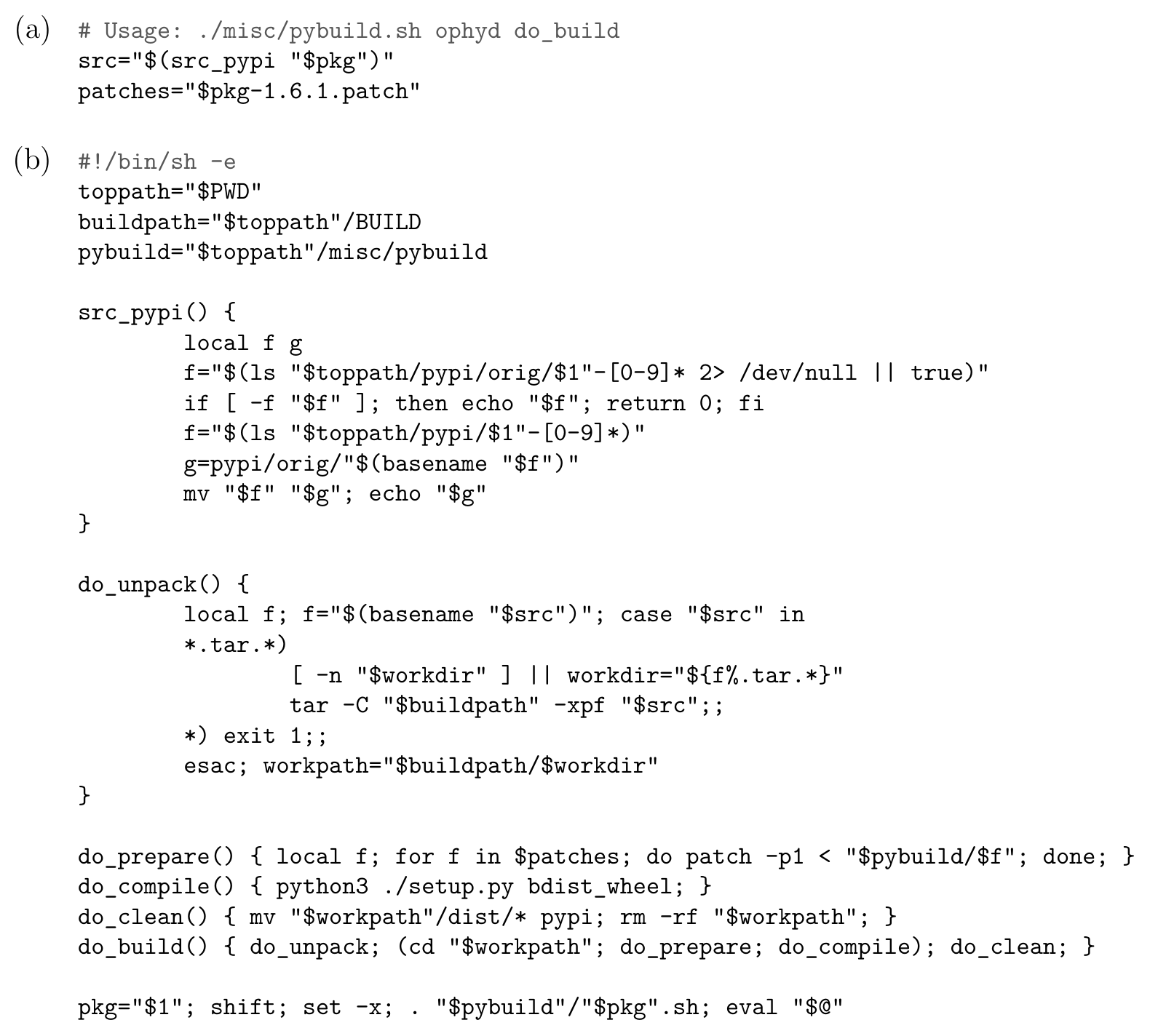}
\caption{%
	(a) \texttt{ophyd.sh}, the definition file of the \emph{ophyd}
	package customised at HEPS; (b) a condensed \texttt{pybuild.sh},
	expected to be executed from the root directory of the build system%
}\label{fig:pybuild-demo}
\end{figure}

Traditional \emph{RPM} packaging does not involve the integrity check of source
archives, which is a common feature in many packaging frameworks, and also
a crucial security feature given the recently increasing threats posed by
\emph{supply chain attacks}.  To compensate for this shortcoming, we maintain
a group of checksum files (\verb|SHA512SUMS| \etc) and compare the checksums
of downloaded source archives against them at the end of the \verb|src_get|
subcommand of \verb|boot.sh|.  Actually, for each input outside of our packaging
system, we have a \emph{chain of trust} (\figref{trust-chain}) to ensure its
integrity: there is \verb|SHA512SUMS-iso| for the CentOS \verb|Everything|
image we use, which ensures every base package contained therein is to an
extent trusted; also in \verb|SHA512SUMS| are the checksums for OpenPGP public
keys of repositories we download third-party packages (including Docker which
is discussed later in this subsection) from, and the signatures of downloaded
packages are checked against the keys at the end of \verb|epel_get|; the base
image which our Docker container images are based upon is also referenced by its
checksum to avoid tampering.  Another issue loosely related to beamline control
is also handled in our packaging system: providing a repository of Python
packages (currently mostly related to \emph{Bluesky}, \cf\ \citealt{allan2019})
for use in beamline experiments, some of them with customised patches and some
of them developed internally at our facility.  We first use the \verb|pypi_get|
subcommand of \verb|boot.sh| to batch-download most of these Python packages
from the Python Package Index (PyPI) and source archives of the several
internally developed packages; after this, we can use the \verb|pybuild|
subcommand to build the customised or internally developed packages (the former
also downloaded as source archives).  The \verb|pybuild| subcommand delegates
most of its tasks to a script, \verb|pybuild.sh| (\figref{pybuild-demo}), which
is essentially a miniature build system for Python packages heavily inspired
by Gentoo \emph{ebuild}.  We also note that given the high fluidity of PyPI
packages, a ``hidden'' advantage of our internal Python package repository
is the relative stability of systems using the packages therein,
assuming our repository is not updated too frequently.


As has been mentioned in \ssref{epkg}, we used to run our packaging system in
virtual machines to avoid interference with actually used machines where
EPICS-related packages have been installed; this also reduced the potential
security risks resulted by password-less \emph{sudo} permission of the builder
user.  Nevertheless, the preparation and usage of virtual machine images for
packaging tasks is still fairly complex and involves quite a number of manual
steps, which reduces reproducibility and consequently reliability; moreover,
file transmission between virtual machines and the host machine is not very
convenient to perform in an automated fashion.  To further increase
reproducibility, we developed the helper script \verb|dboot.sh|
(\figref{boot-usage}(b)) to create suitable Docker container images and
perform packaging tasks in containers instantiated from these images; built
\emph{RPM} and Python packages are respectively put into the \verb|RPMS|
and \verb|pypi| directories (\figref{builder-skel}) automatically, by using
the ``mounting'' feature of Docker.  To save time used in installing common
build-time dependencies, we distinguish between \verb|base| and non-\verb|base|
container images: the former used to build basic packages like EPICS base
and \emph{synApps}, with preparation tasks (like setting up password-less
\verb|sudo|) done on image creation time; the latter based on the former but
with the basic packages pre-installed (and therefore needs to be recreated when
these packages are updated).  A common practice in software engineering is
\emph{continuous integration} (CI), where updates to the code repository
trigger rebuild of the changed parts automatically.  We currently do not have
a fully automatic CI workflow, due to the complexity in managing source archives
as well as in rebuilding dependencies and container images; however, we also
find the current script-assisted rebuilding procedure sufficiently succinct
and efficient that it can be regarded as some kind of ``slightly manual'' CI.

\section{Minimisation of configuration}
\subsection{Reusable modular IOC executables}\label{ssec:app}

Most time in EPICS-based beamline control are spent on maintaining
IOC (``input/output controllers'') applications which translate various
device-specific control protocols into the Channel Access (CA) protocol
exposed by EPICS.  IOC applications are mainly composed of IOC executables
(based on underlying support libraries) and data files (most importantly
\verb|.dbd| files, \verb|.db| files and \verb|.cmd| files).  From a slightly
academic viewpoint, IOC executables are interpreters that interpret \verb|.cmd|
files, which most importantly tell interpreters to load specified \verb|.dbd|
files and \verb|.db| files.  Upon closer observation, it is not difficult
to find that for IOC applications that handles the same type of devices, the
source files of the IOC executables and \verb|.dbd| files are mostly unchanged:
although the filename of the executable and the \verb|.dbd| file may change,
these factors do not lead to any substantial change in the behaviours of the
IOC application.  In other words, the exposed interface of an IOC application
is mostly affected by the interpreted files excluding \verb|.dbd| files; we
may consider \verb|.dbd| files as an external yet closely coupled part of IOC
executables, and other data files as configuration files in general.  For many
support modules, it is common to provide shared \verb|.db| files, and let
users compose them into more complex interfaces by using \verb|.substitutions|
files; we regard this kind of shared data files as configuration fragments
provided by upstream that can be composed or customised by the user.  From the
above, it may be concluded that for IOC applications dealing with at most one
type of device, we may provide reusable IOC executables (including \verb|.dbd|
files and other shared data files) that can be fed with different configuration
files to act as different application instances.  We will show in the end of
this subsection that application instances for different devices, seen as
modules, can be composed to implement the same requirements done by at least a
big fraction of previous multi-device IOC applications; thanks to this change,
we can minimise the need for users to build IOC applications by themselves,
which greatly reduces the need for systems like \emph{Sumo} \cite{franksen2019}.
The applications instances that replace a multi-device application can still
be run on the same computer, and be started/stopped as a group; since they
would then communicate through the local loopback interface which can only
be disrupted in the case of very serious errors, the splitting does not hurt
reliability.  By properly organising configuration files for different instances
of the same IOC executable (\cf\ \ssref{cfg}), conflicts between different
instances can be easily avoided, so we do not need to use mechanisms like
Docker to isolate different IOC applications \cite{derbenev2019}, which
simplifies deployment and saves both memory and disk space.  Another point
worth mentioning is that the idea of \emph{reusable modular IOC executables}
seems more natural and easier to understand for us than the \emph{require}
mechanism \cite{psi2015}, which loads support libraries on demand in
\verb|.cmd| files and requires much more patching in the packaging
procedure for every EPICS support module involved.

\begin{figure}[htbp]\centering
\includegraphics[width = 0.8\textwidth]{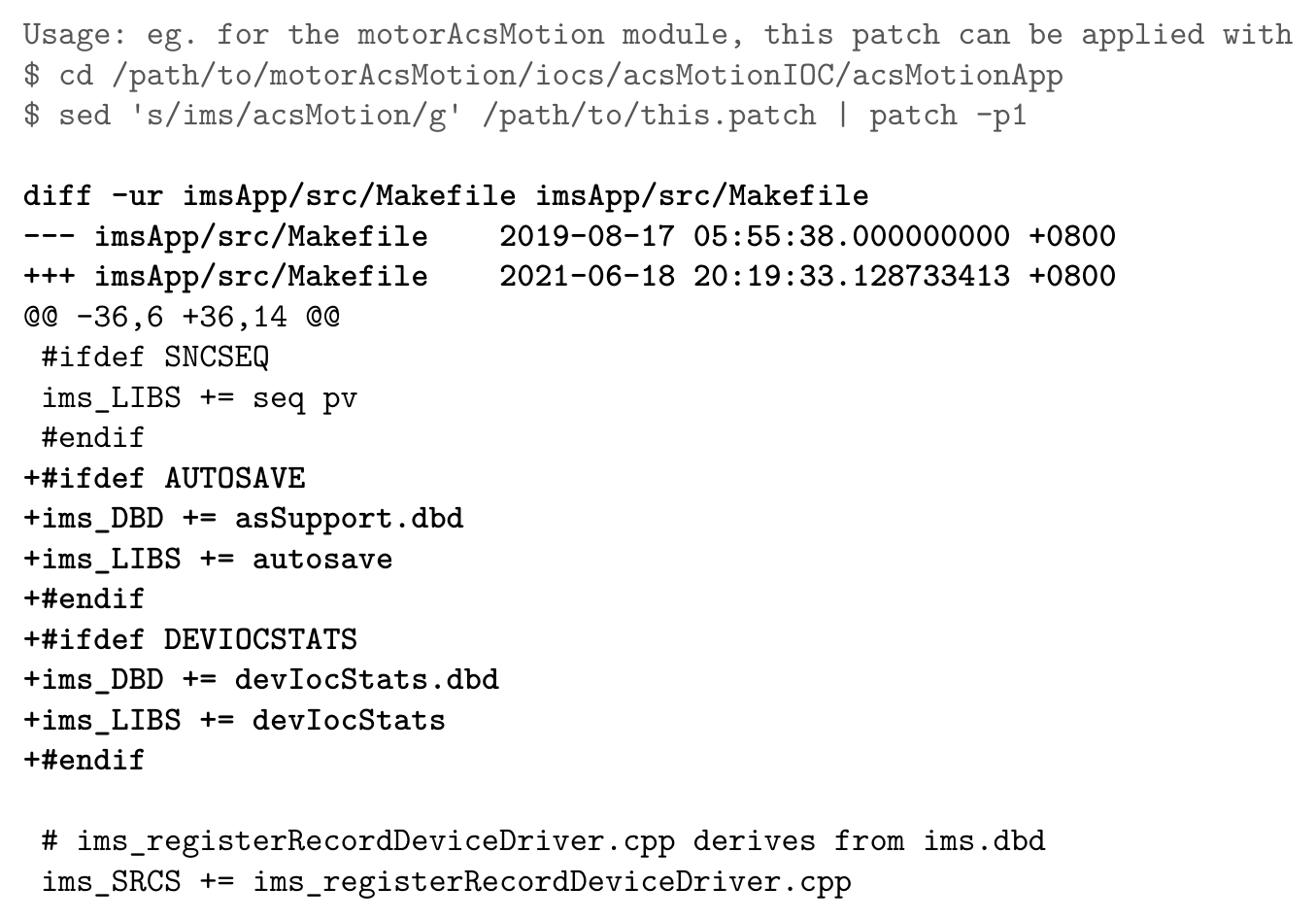}
\caption{%
	A ``template'' used to batch-add support for
	\emph{iocStats} and \emph{autosave} to motor modules%
}\label{fig:motor-patch}
\end{figure}

For reusable modular IOC executables to be provided, they must be built first;
for some EPICS modules this is done by default, or is disabled but can be easily
enabled.  For other modules which we want reusable executables of, we need
to build them by ourselves; for various purposes, we also add features like
\emph{iocStats} (IOC ``health'' monitoring) and \emph{autosave} (automatic state
saving and restore) to most IOC executables.  We do these by patching the build
system (usually \texttt{configure/RELEASE} and \texttt{xxxApp/src/Makefile} in a
module) before actually running it; the centralised management of packaging code
(\cf\ \ssref{builder}) also helps to review patches and compare the patches of
modules that are similar in packaging, both of which prove to be very helpful
in quality assurance.  The build systems of modules that belong to a common
type (most importantly motors and area detectors) often only differ in a few
highly ``templated'' locations.  This similarity can sometimes be exploited to
simplify the patching procedure above: since the patches for these modules would
also be templated, we may provide a template (\figref{motor-patch}) and generate
the patches on the fly for them based on this template.  For this reason, we
prefer to make the build systems of internally developed \emph{motor} and
\emph{areaDetector} modules similar to their counterparts in \emph{synApps},
instead of using the more ``standardised'' (and much more bloated) build system
generated by traditional tools like \verb|makeBaseApp.pl|; to generalise a
little, we have the policy for fundamentally similar code in different places
that \emph{if they cannot yet be abstracted, at least make them templated}.
A situation where we further exploit the similarity between EPICS build systems
is building generic extra modules: we find that although the build systems of
these EPICS modules (\eg\ \emph{Keithley\_648x} for Keithley 6485/7 picoammeters
and \emph{s7nodave} for Siemens PLCs) do not look immediately similar, they can
usually be replaced by a ``standard'' counterpart with only minimal changes in
some ``template parameters''.  For this reason, when packaging \emph{synApps},
we provide standard \verb|configure| directory and \verb|Makefile|s that would
appear in \verb|xxxApp|, \verb|iocBoot| and \verb|iocxxx|, respectively at
\verb|configure|, \verb|app.mk|, \verb|iocBoot.mk| and \verb|ioc.mk| in the
\texttt{/opt/epics/support/utils} directory, and have successfully used them
to replace their module-specific counterparts.

When we talked above reusing IOC executables above, what we meant to reuse
was not only those that deal with specific devices, but also those that do
not directly communicate with any real device.  They may control other devices
indirectly by communicating with other applications through the CA protocol,
like the \emph{optics} (monochromators and similar utilities) and \emph{sscan}
(interlocked action of motors and detectors, functionally like \emph{Bluesky})
modules; they may also provide PVs (``process variables'' in EPICS) which other
applications may read or write through the CA protocol.  The behaviours of
some applications among them are completely determined by the \verb|.db| files
supplied, to the point that their behaviour remain intact if they are instead
fed to the intuitively named \emph{softIoc} executable instead; this is how we
implement these applications, so that we do not even need to write \verb|.cmd|
files for them.  However, given our requirement of status monitoring mentioned
above, we also need to add the \emph{iocStats} feature to our \emph{softIoc}
executable (\cf\ \figref{run-script}(a)), which results in a chicken-and-egg
situation: \emph{synApps} depends on EPICS base which provides \emph{softIoc},
while \emph{softIoc} needs to be added with support for \emph{iocStats} which
is from \emph{synApps}.  We break this dependency loop by disabling the building
of \emph{softIoc} in EPICS base, lifting the former's source code from the
latter, using the templates in \texttt{/opt/epics/support/utils} to form the
build system for a standalone \emph{softIoc}, and finally building it with
\emph{iocStats} support added.  Another dependency loop is that the \emph{sscan}
IOC executable depends on the \emph{calc} module, which in turn depends on the
\emph{sscan} library; we solve it in a similar way -- first building libraries
in the two modules only, and then building the executables separately.  We also
note that in order to make references to EPICS modules (including executables
and libraries) stable across updates, we remove the version tags from the
directory names of all \emph{synApps} modules (like the \verb|-R4-36| from
\verb|asyn-R4-36|, \cf\ also \figref{run-script}); this does not prevent
the user from querying the version numbers, which can still be found
in \texttt{/opt/epics/support/assemble\string_synApps.sh}.  And to make
OPIs (``operator interfaces'', GUIs for IOCs) easier to find, we create
symbolic links to all OPI files under \texttt{/opt/epics/support} in
\texttt{/opt/epics/resources}: links to \verb|.edl| files for \emph{EDM} in
the \verb|edl| subdirectory, links to \verb|.adl| files for \emph{MEDM} in the
\verb|adl| subdirectory, \etc.  Search paths like \verb|${EPICS_DISPLAY_PATH}|
are also set to these resource subdirectories, so that users do not even need
to type the directory names for OPI files; in case of ``missing'' resource
files (\eg\ \verb|kohzu.gif| referenced by \verb|kohzuGraphic.adl| from
the \emph{optics} module, expected to be in the same directory),
the user can also easily find the real path of OPI files
by looking at where the symbolic links point to.

\begin{figure}[htbp]\centering
\includegraphics[width = 0.5\textwidth]{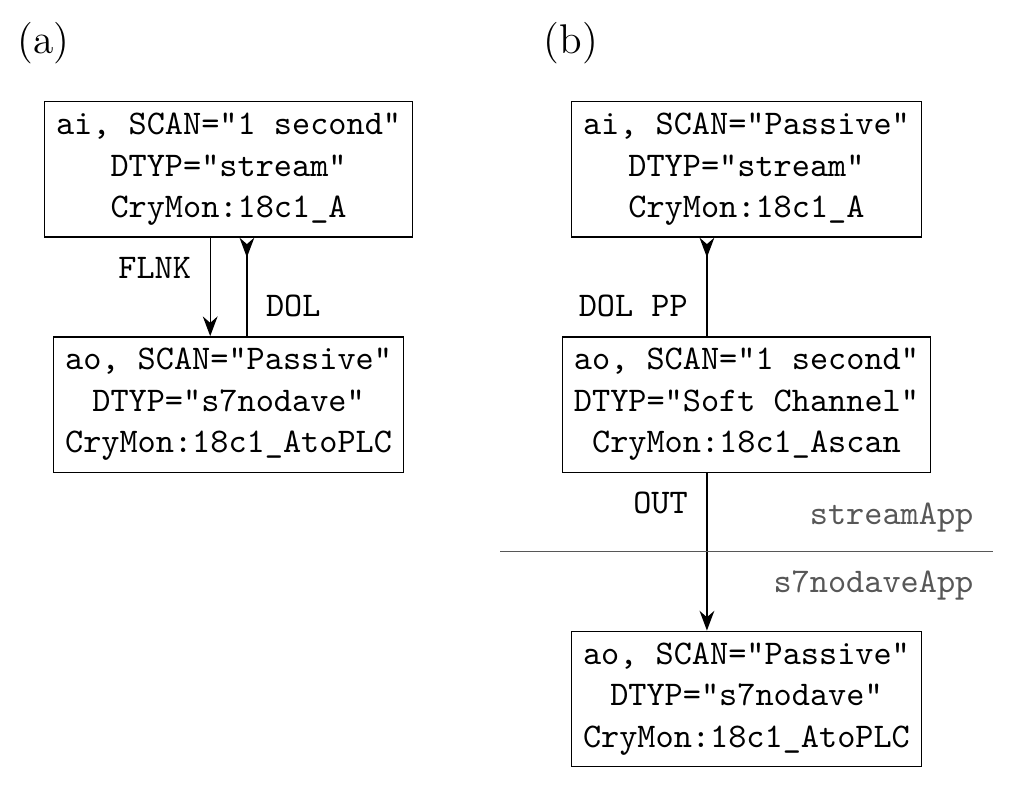}
\caption{%
	A requirement implemented with (a) multi-device and (b) single-device
	IOC applications; for reliability reasons, the latter should be
	run on a same computer, and be started/stopped as a group%
}\label{fig:link-refactor}
\end{figure}

Interlocked actions between devices are a well-known strength of EPICS,
and we provide full compatibility with self-building of multi-device IOC
applications: the libraries and ancillary files are provided as usual, with
reusable modular IOC executables just as an added bonus.  However, we also find
what is done by at least a big fraction of previous multi-device applications
can also be done by the composition of their single-device counterparts; the
first examples are \emph{optics} and \emph{sscan}, which in our experiments work
well whether the devices involved are controlled by the same IOC application or
not.  From a theoretical perspective, a multi-device application can be split
into multiple single-device applications as long as the main modules involved
(like \emph{sscan} and the motor/detector modules it is supposed to control)
do not actually assume each other were in the application, or in EPICS terms
they do not mandate that they communicated through DB links instead of CA links.
For some applications, even if this is not true, the link relation between the
modules can be refactored to eliminate the demand for DB links; this certainly
needs to be analysed on a case-by-case basis, but here we give a relatively
simple real-world example that is nevertheless fundamentally similar to many
other IOC applications in production, which is why we believe many multi-device
applications can actually be split up.  For a certain application scenario, we
need to poll temperatures from a group of Cryo-con 18C monitor channels, and
write them to a group of memory addresses of Siemens PLCs; our previous and
current solutions are shown in \figref{link-refactor}.  For each pair of
temperature channel and PLC address, we use a pair of EPICS analogue-input
and analogue-output PVs based on \emph{StreamDevice} (\cf\ \ssref{cfg}) and
\emph{s7nodave} respectively.  In the previous solution, the \emph{StreamDevice}
PV periodically updates (``scans'') from the temperature channel, and after
every update uses a ``forward link'' to trigger update of the \emph{s7nodave}
PV, which pulls the temperature from the \emph{StreamDevice} PV and writes
it to the specified memory address.  According to the EPICS documentation
\cite{kraimer2018}, a CA forward link only works as expected if it explicitly
points to the \verb|.PROC| ``field'' of the destination PV (\eg\ %
\texttt{CryMon:18c1\string_AtoPLC.PROC}), so we can eliminate the demand
for DB links by batch-adding the \verb|.PROC| suffix to the forward links.
For multiple reasons, we actually use a slightly more complex solution,
with ``soft channel''-based analogue-out PVs (supported by EPICS base) as
intermediates for temperature values; the ``soft channel'' PVs are scanned
instead of the \emph{StreamDevice} PVs, and the latter are only passively
updated (using the ``process passive'' option) before the \emph{s7nodave} PVs.

\subsection{Configuration minimisation and automated deployment}\label{ssec:cfg}

An issue with introducing package management into EPICS is the management
of configuration files.  The first problem is that if the user tunes the
behaviours of IOC applications by directly changing packaged files, these
changes would get lost when the corresponding package is updated; although
package managers usually support the declaration of certain packaged files (like
all files under \verb|/etc|) as ``configuration'', which helps to prevents the
loss of configuration changes, there are often too many customisable files
in EPICS to be batch-declared as configuration.  The second problem is that
since packaged files are typically installed as \verb|root|-owned (\cf\ %
\ssref{epkg}), we often would need to use \verb|root| privilege to change them,
with implications in security and even convenience: \eg\ with reusable IOC
executables (\cf\ \ssref{app}), the same executable may be used by multiple
users for different applications, and one user may accidentally overwrite
configuration files that actually belong to another user.  Both problems can
be easily solved if the configurations files by each user are kept inside
the user's own home directory; this is a common practice outside the EPICS
ecosystem, and is in essence analogous to the separation of packaging code
and source archives (\cf\ \ssref{builder}) in that both distinguish between
upstream and downstream (upstream developers \vs\ packagers, packagers \vs\ %
users) in order to reduce the amount of code each party needs to maintain.
As has been mentioned just now and in \ssref{builder}, this also helps with
software update and migration: a very nice side effect of the separation
of configurations is that if somehow we find a clean way to get rid of the
packaging-unfriendly \emph{synApps} directory layout (\cf\ \ssref{epkg}),
users will be able to migrate to the new layout relatively easily by
batch-changing all path references in configurations files accordingly.

\begin{figure}[htbp]\centering
\includegraphics[width = 0.86\textwidth]{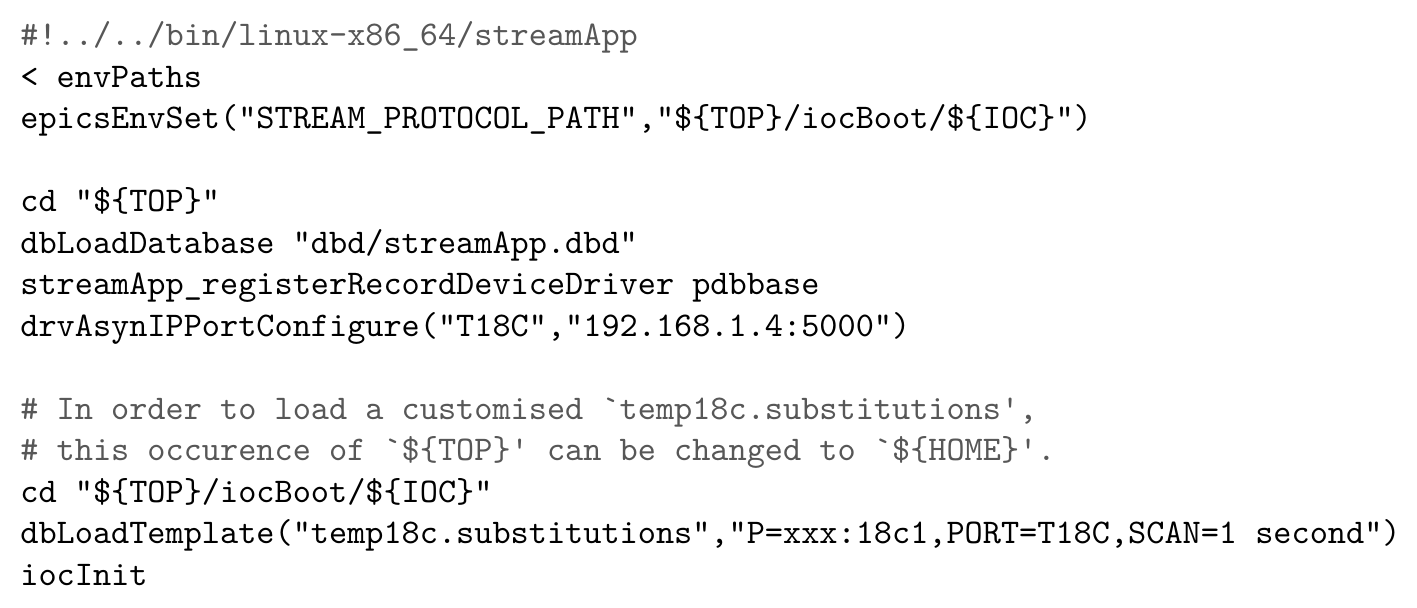}
\caption{Default \texttt{.cmd} file for Cryo-con 18C temperature monitors}
\label{fig:cmd-demo}
\end{figure}

\begin{figure}[htbp]\centering
\includegraphics[width = 0.86\textwidth]{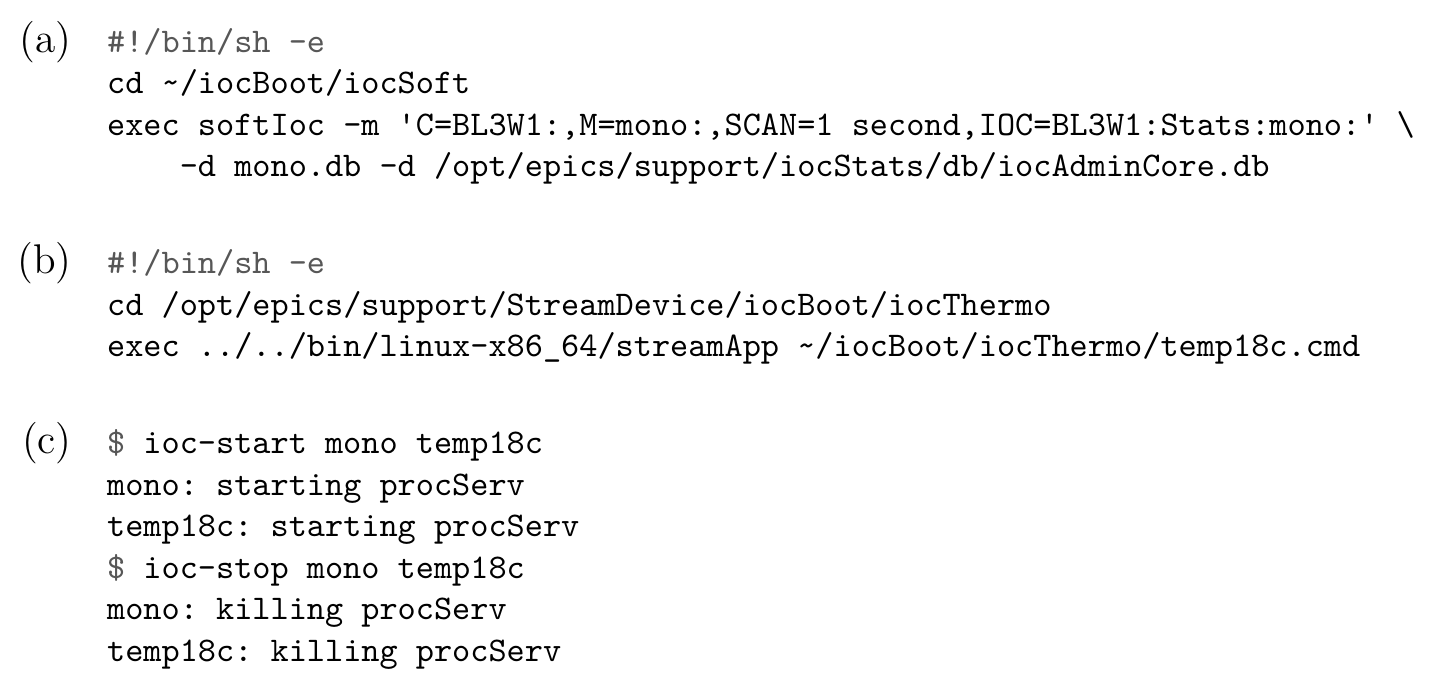}
\caption{%
	Run scripts in \texttt{\$\string{HOME\string}/iocBoot/service} in a
	production environment for (a) a LabView-based monochromator gatewayed to
	EPICS with \emph{CA Lab} (\texttt{run-mono.sh}) and (b) some Cryo-con 18C
	temperature monitors (\texttt{run-temp18c.sh}), with (c) example invocations
	of utilities that start and then stop corresponding applications%
}\label{fig:run-script}
\end{figure}

At our facility, we have formed what we call \emph{the \texttt{\~{}/iocBoot}
convention}: putting configurations corresponding to an EPICS module (\eg\ %
\emph{motorIms} for MDrive motors) into a suitable subdirectory (\eg\ %
\emph{iocIms}) of \verb|${HOME}/iocBoot| named after the subdirectory of
\verb|iocBoot| where the configuration files are derived from.  This is
chosen because \verb|.cmd| files (\figref{cmd-demo}) usually change to the
``\verb|${TOP}/iocBoot/${IOC}|'' directory before loading their default
configurations, so we can easily adapt them by changing the references into
``\verb|${HOME}/iocBoot/${IOC}|'' where appropriate.  For modules where the
\verb|${IOC}| subdirectories are named too generally (\eg\ \verb|iocAny| from
\emph{optics}), the packager renames them into more distinguishable names
(\eg\ \verb|iocOptics|); for packages where \verb|iocBoot| is not provided
or the contents are not very good examples, the packager provides self-made
examples that can be easily customised.  The \emph{StreamDevice} module is
widely used in EPICS for talking to simple devices in a request/reply fashion,
where developers mainly need to provide \verb|.proto| ``protocol files'' and
corresponding \verb|.db| files; we have a dedicated package that collects
these files, and groups them into directories like \verb|iocCounter| and
\verb|iocThermo| according to the device type (\cf\ \figref{run-script}(b)).
\emph{softIoc} is a special case where the \verb|iocBoot| directory is not
strictly needed, but in the name of uniformity all \verb|.db| files for
\emph{softIoc} are put in \verb|${HOME}/iocBoot/iocSoft|.  Furthermore, all
\verb|.cmd| files are patched to store \emph{autosave} state files in
\verb|${HOME}/iocBoot/autosave/${IOC}|; nevertheless, to prevent IOC
executables from outputting warning messages over and over when the specified
directories do not yet exist, the saving and loading of state files are disabled
by default.  We use the \verb|${HOME}/iocBoot/service| directory to store
``run scripts'' (\figref{run-script}) that when executed run corresponding
applications in the foreground; utilities based on \emph{procServ} are also
provided to start/stop specified application(s) in the background, and
we are developing \emph{procServControl}-based mechanisms to provide GUIs
for centralised status management.  With all these elements combined, we
are able to implement multi-application setups, where the hundreds of IOC
applications needed for a beamline (excluding \emph{areaDetector} and other
applications that are resource-intensive, as well as VME-based applications)
can be accommodated on just a few computers with very modest hardware.  We
find this much easier to implement and maintain than alternatives, \eg\ %
\cite{konrad2019} which uses much more abstraction based on \emph{Puppet}.


System administration is closely related to the concept of ``configuration
management''; to us, configuration is the entire procedure of composing hardware
and software, whether off-the-shelf or self-developed, into systems that can
readily work in production.  In the above, we have shown the basic aspects of
our efforts to minimise the complexity in configuring software on a single
computer for comfortable EPICS-based beamline control; in addition to these,
we also further reduce the complexity of said configuration by simplifying
configuration files using various mechanisms, whether based on EPICS (\eg\ %
\verb|.substitutions| files) or not (mainly code generators, provided the
workload saved by them significantly outweighs the workload to maintain them).
To also simplify hardware configuration, we have written a \emph{software/%
hardware handbook} to give simple yet reproducible instructions that can be
followed to configure beamline devices and corresponding software into basic
working states.  On a larger scale, we are developing a backup system where the
actual configuration of computers (whether running IOCs, OPIs or software like
\emph{Bluesky}) are automatically synchronised to a central backup service,
which in turn distributes the computer configurations on a beamline to every
computer on this beamline, so that the configuration of a computer can be
replicated onto a backup computer in case of hardware error.  Git-based version
control will be applied to the backups, so that we can revisit recent changes
in case misconfiguration is suspected; in order to minimise downtime spent on
replicating configuration, automation mechanisms will be developed for the
replication procedure, and backup computers will be preloaded with basic
software after procurement if applicable.  In addition to regular computers,
we also plan to extend the backup system to other programmable hardware, like
PLCs and network switches, that accept automated and structured configuration
input, so that their configuration procedures can also be simplified.

From our definition of ``configuration'' above, we can see that reducing the
amount of configuration for a single beamline device is reducing the amount of
device-specific workload, and that reducing the amount of configuration of an
entire beamline is reducing the amount of beamline-specific workload.  Therefore
in order to maximise the scalability of beamline control, we are developing
a group of \emph{comprehensive beamline services} (CBS), which covers most
aspects in beamline control that are shared between beamlines.  A documentation
library service will be provided, which gives users access to not only the
software/hardware handbook (also including a recommended list of hardware for
various use on a beamline) and the operation manuals of individual beamline
devices, but also training materials (which would ``configure'' new users for
various aspects of the facility) and other useful documentation.  Centralised
network services will be provided, including the backup system and the
documentation library mentioned above, our package repositories (\cf\ %
\ssref{builder}) and other services like an EPICS PV archiver/alarm service
and CA gateways to forward information from \eg\ the accelerator.  For
reliability reasons, our control network (where control information is
transmitted) will be isolated from the data-transfer network (where bulk data
produced by area detectors and data produced by software like \emph{Bluesky}
are transmitted).  Beamlines cannot directly communicate with each other, and
instead can only communicate with the CBS; transmission of outside information,
like NTP information and PVs from the accelerator, is done by gateway services
in the CBS.  A beamline development laboratory (BDL) will be provided, where
beamline devices can be tested without interfering with the production
environment; the BDL is like a small beamline in terms of
networking, which the CBS is formally a part of.

\begin{figure}[htbp]\centering
\includegraphics[width = 0.9\textwidth]{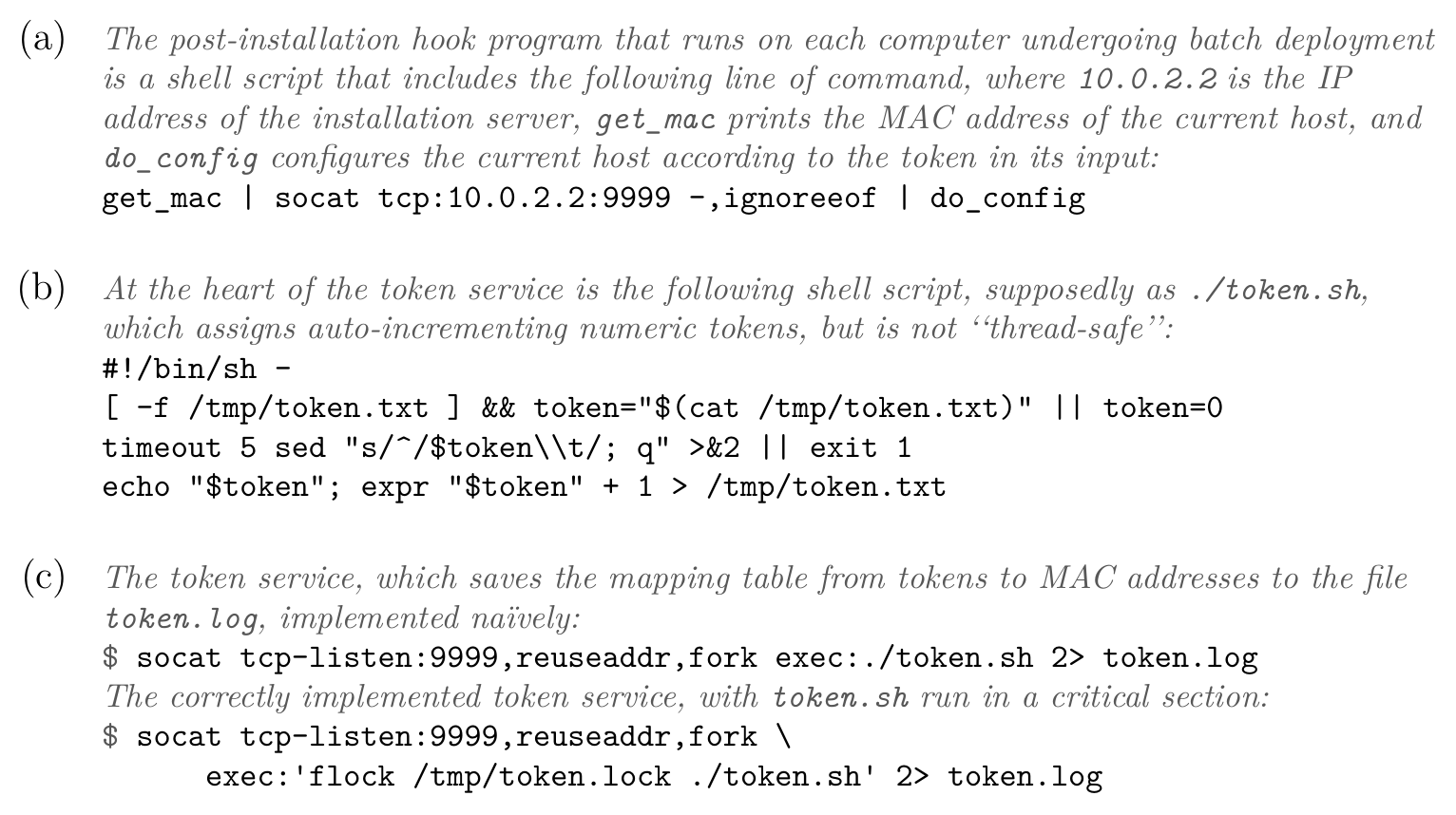}
\caption{Implementation of a token service}\label{fig:token-service}
\end{figure}

Another prominent use of the BDL is batch deployment, where basic software is
installed on new hardware before the latter is moved to the beamlines or stored
as backup hardware.  In the end of this paper, we use the batch installation of
operating systems onto new computers as an example for how certain requirements
in system administration can be implemented in surprisingly simple ways, when
we compose simple mechanisms and utilities according to careful analyses of
the nature of the problems.  Regular installation of operation systems is
often performed with CDs/DVDs or USB flash drives, which are exclusive media
that cannot be accessed by multiple computers simultaneously; thus for batch
deployment, network-based installation (now almost universally based on PXE) is
preferred, which is only limited by the network bandwidth of the installation
environment and perhaps disk performance of the installation server.  In order
to facilitate large-scale deployment, network installers of Linux distributions
often support some kind of templated unsupervised installation mechanism, like
\emph{Kickstart} of CentOS; many of them, including \emph{Kickstart}, also
support some kind of mechanism that allows for post-installation execution of
specified programs (``hooks''), so we can automatically preload basic software
during unsupervised installation.  However, for some requirements, we also need
differentiation between computers, \eg\ installation of systems onto a group of
servers for EPICS PV archiving with automated configuration of hostnames and
related settings.  This can be done if we set up a service on the installation
server that automatically assigns tokens (which the hostnames will be based on)
according to certain rules, and let the hook program obtain the token from the
service; using the \emph{socat} utility and a little Shell scripting, this can
be done in just a few lines of code (\figref{token-service}).  The na\"ive
implementation of the service suffers from an obvious race condition when
multiple clients ask for tokens at the same time; this can be solved
if we execute the service script in a critical section, which can be
easily implemented by the \emph{flock} utility.

\section{Conclusion}

At HEPS, we currently use \emph{RPM} for packaging beamline control software,
and deliberately use EPICS 7 like EPICS 3, with most modules based on
\emph{synApps}.  We preserve the \emph{synApps} directory layout in our
packaging system, which however results in potential file conflicts and
permission issues; we solve these problems using a builder user with
password-less \emph{sudo} permission in Docker containers.  The packaging
system, including the packaging code and the Docker wrapper, is managed
centrally and properly abstracted to minimise complexity.  Both self-built
packages and some very useful third-party packages are provided in our internal
\emph{RPM} repository, and we also provide a similar Python package repository
for internal use; all external inputs involved in the creation of both
repositories are checked against a chain of trust to avoid tampering.
We use reusable modular EPICS IOC executables, built with support for
\emph{iocStats} and \emph{autosave}, to minimise the need for self-built
multi-device applications, and facilitate the use of these executables by
providing easy access to resources like OPIs and example configuration files.
Full compatibility with multi-device applications is kept, and we also find
at least a big fraction of them can be replaced by the composition of their
single-device counterparts.  We have formed the \verb|~/iocBoot| convention
to separate each user's IOC configuration files from the default configurations
provided by \emph{RPM} packages, and provide utilities that help to implement
maintainable multi-IOC setups for beamlines.  Rigorous efforts are being made
under the umbrella project of comprehensive beamline services to further
simplify configuration management on multiple scales: beamline devices and
related software on individual computers, all computers and other programmable
hardware on an entire beamline, as well as all beamlines at HEPS.

\subsection*{Acknowledgements}

We would like to thank Yu-Jun Zhang and Wei Xu for the first adoption
of the \verb|~/iocBoot| convention and their continuous feedback.

\bibliography{art3}
\end{document}